\documentclass[conference,a4paper]{IEEEtran}
\IEEEoverridecommandlockouts

\usepackage[hidelinks]{hyperref}
\usepackage[cmex10]{amsmath}
\usepackage{amssymb,amsfonts}
\usepackage{dblfloatfix}

\usepackage[ruled,vlined]{algorithm2e}
\usepackage{graphicx}
\graphicspath{{Figures/PDF/}{Figures/PNG/}}

\usepackage{booktabs}
\usepackage{siunitx}
\usepackage[numbers,compress]{natbib}
\usepackage{texnames}
\usepackage{bm,bbm}
\usepackage{orcidlink}
\usepackage{graphicx}   
\usepackage{xcolor}     
\usepackage{colortbl}   
\usepackage{diagbox} 
\usepackage{threeparttable} 

\begin{document}

\title{\uppercase{Improved VMD Based Remote Heartbeat Estimation Utilizing 60GHz mmWave Radar}}

\author{
        \hspace{-3.5em}\IEEEauthorblockN{Boyuan Gu}
	\IEEEauthorblockA{\hspace{-4em}\textit{University of Electronic Science and Technology of China}\\
		\hspace{-4em}2006 Xiyuan Ave. Chengdu, China\\
		\hspace{-4em}guboyuan79@gmail.com}
	\and
	\IEEEauthorblockN{Yanhui Yang}
	\IEEEauthorblockA{\textit{University of Electronic Science and Technology of China}\\
		2006 Xiyuan Ave. Chengdu, China\\
		2022190502027@std.uestc.edu.cn}
	\and
	\IEEEauthorblockN{Siyu You}
	\IEEEauthorblockA{\textit{University of Electronic Science and Technology of China}\\
		2006 Xiyuan Ave. Chengdu, China\\
		2022190502023@std.uestc.edu.cn}
        \and
        \IEEEauthorblockN{Haiyang Sun}
	\IEEEauthorblockA{\textit{University of Electronic Science and Technology of China}\\
		2006 Xiyuan Ave. Chengdu, China\\
		15845945009@163.com}
         \and
        \IEEEauthorblockN{Jiahui Sun}
	\IEEEauthorblockA{\textit{University of Electronic Science and Technology of China}\\
		2006 Xiyuan Ave. Chengdu, China\\
		3423349769@qq.com}
        \and
        \IEEEauthorblockN{\hspace{4em}Shisheng Guo*}
	\IEEEauthorblockA{\textit{\hspace{3.5em}University of Electronic Science and Technology of China}\\
		\hspace{3.5em}2006 Xiyuan Ave. Chengdu, China\\
		\hspace{3.5em}ssguo@uestc.edu.cn}
}

\maketitle
\begin{abstract}
	This study introduces an improved VMD based signal decomposition methodology for non-contact heartbeat estimation using millimeterwave (mmWave) radar. Specifically, we first analyze the signal model of the mmWave radar system. The Variational Mode Decomposition (VMD) integrated with the Newton-Raphson-based optimizer (NRBO) algorithm are sequentially utilized for cardiac mechanic signal (CMS) reconstruction.  The estimation accuracy is enhanced by adaptively optimizing the VMD parameters including intrinsic mode functions (IMFs) and penalty factor ($\alpha$). Eventually, the experimental results of 18 subjects validate the effectiveness of the proposed method by comparing with three commonly used baselines.
\end{abstract}


\section{Introduction}

Heartbeat estimation is a critical component in medical diagnostics and health management. Conventional methods for heartbeat estimation primarily rely on contact-based electrocardiogram (ECG) signals, which can be inconvenient and impractical in certain scenarios, especially long-term sleep monitoring [1]. In recent years, radar-based technology Millimeterwave (mmWave) radar has emerged as a promising noncontact technology for vital signal detection, owing to its high sensitivity, robust penetration, and adaptability to diverse environments [2]. These characteristics make mmWave radar well-suited for applications requiring continuous and non-invasive monitoring, providing an alternative to traditional ECG-based methods.

Xu \textit{et al.} [3] introduced an application of Variational Mode Decomposition (VMD) for cardiac cycle signal monitoring in complex environment. Nevertheless, it is difficult to solve the challenges posed by noise and signal overlap. Du \textit{et al.} [4] proposed an optimized GA-VMD approach, which integrates genetic algorithms to improve parameter selection. This method has achieved higher accuracy in separating heartbeat compared to standard VMD methods. However, due to the iterative nature of genetic algorithms, it is susceptible to local optima during parameter optimization [5].

To further advance the accuracy of noncontact heartbeat estimation, we introduce a novel method that integrates a Newton-Raphson-based optimizer (NRBO) with VMD, focusing on optimizing the penalty factor and identifying the most relevant intrinsic mode functions (IMFs) and penalty factor $\alpha$ to further improve the precision of heart rate estimation. NRBO dynamically adjust VMD parameters to enhance signal decomposition, effectively isolating cardiac components. The effectiveness of the proposed model is evaluated through experimental results involving cardiac mechanical signals (CMS) from 18 participants, captured using a 60 GHz millimeter-wave radar system.

\section{Methodology}

\subsection{Signal Model of mmWave Radar}

\begin{figure*}[htbp]
\centering
\includegraphics[width=0.8\textwidth]{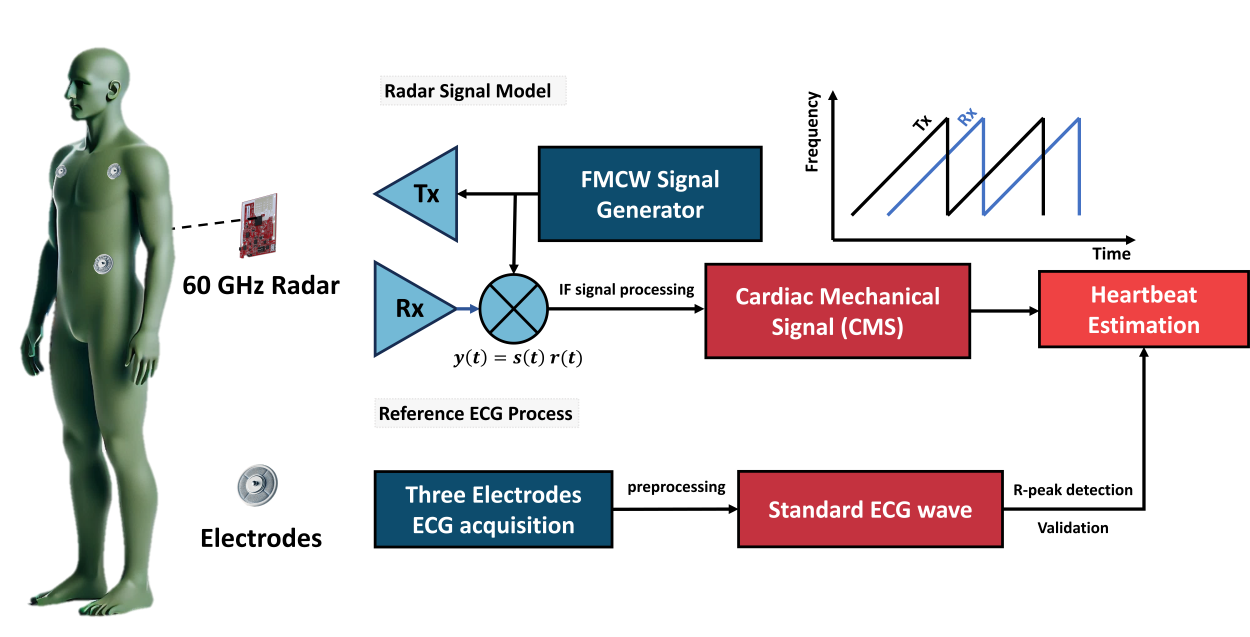}
\caption{Process of experimental signal acquisition.}
\label{fig1}
\end{figure*}

FMCW radar, featuring as linearly modified frequency region within each chirp, is capable of acquiring the signals with weak amplitude by extracting the phase signal. Therefore, FMCW radar is an important modality applied to non-contact physiological signal measurement. The process of experimental signal acquisition is illustrated in Fig. \ref{fig1}.

Plotting the transmitted and received signals in the frequency domain for two chirps, the $s(t)$ and $r(t)$ signals can be expressed in plural form.

Based on Fig. 1, the transmitter signal of radar can be expressed as:

\begin{equation}
    s(t)=e^{j(2\pi f_c t+\pi \frac{B}{T}T^2)}
\end{equation}
    \begin{equation}
        r\left(t\right)=e^{j\left(2\pi f_c\left(t-t_d\right)+\pi\frac{B}{T}\left(t-t_d\right)^2\right)}
    \end{equation}
where $f_c$ is the carrier frequency of the transmitter signal; $B$ is the bandwidth of FMCW; $T$ is the sweep time of one chirp and $t_d$ is the time delay of echo signal.
After passing through the mixer, the signal $y(t)$ can be derived from the multiplication of signal $s(t)$ and $r(t)$:

\begin{equation}
    y\left(t\right)=s\left(t\right)r\left(t\right)=e^{j\left(4\pi\frac{BR}{cT}+\frac{4\pi}{\lambda}R\right)}
\end{equation}
where $R$ is the target distance.
Based on Formula (3), the target distance can be detected from the phase of the IF signal:

\begin{equation}
    \delta\phi_y=4\pi\frac{\delta R}{\lambda}
\end{equation}

Through [5], the tiny displacement caused by the human heartbeat is between 0.2 to 0.5 mm, and by human respiration is between 1 to 12 mm, which is below the minimum frequency resolution. Therefore, the distance dimension must be estimated by the phase difference in the heartbeat measurement.
For $f_c=60$ GHz , the wavelength $\lambda=5$ mm. It can be derived from Formula (4) that if a displacement $\delta R=1$mm is detected, the phase change will be $\delta \phi = 2.51$radians , which is detectable.
\subsection{ Radar Signal Processing Based on Improved VMD}
\subsubsection{VMD Parameters}

VMD is an adaptive signal decomposition method which decomposes the given signal into a set of IMFs by solving an optimization problem illustrated in Equation (5).

\begin{equation}
    \min_{\{u_k\}, \{\omega_k\}} \sum_{k=1}^K \left\| \partial_t \left[ \left( \delta(t) + j\frac{1}{\pi t} \right) \ast u_k(t) \right] e^{-j\omega_k t} \right\|_2^2
\end{equation}
where $K$ is the number of the IMFs, $\omega_k$ is the center frequency of the $k$th IMFs. The bandwidth of each IMFs is related to the penalty factor $\alpha$.

In the signal processing of the cardiac mechanical signal (CMS), we select and reconstruct the IMFs within frequencies between 0.5Hz to 2Hz, indicating the heart rate of 30 BPM to 120 BPM which is the range of the vast majority of people.  

\subsubsection{Parameter Optimized by Newton-Raphson-based Optimizer (NRBO)}

The performance of VMD is closely related to the IMF number $K$ and the penalty factor $\alpha$. This requires VMD to adjust its parameters in order to achieve optimal signal decomposition and reconstruction when dealing with CMS from different individuals. A stable performance can be achieved if applying an optimization algorithm to automatically adapt the parameters regarding to the 
different characteristics of CMS.

Due to the considerable temporal consistency of CMS, the degree of uniformity of the signal over the time series is highly correlated with its superiority. Thus, the Sample Entropy [6] is chosen as the objective function for the optimization process.

The NRBO algorithm simulates genetic, cooperative, and competitive behaviors thorugh combining the Newton-Raphson Search Rule and the Trap Avoidance Operator. It demonstrates strong performance in convergence rate and the avoidance of local optimal.
The whole optimization process is depicted in Algorithm 1.
\begin{algorithm*}
\caption{NRBO for Optimizing VMD Parameters}
\label{alg:NRBO}
\KwIn{$N$: Population size, $max\_iter$: Maximum iterations.}
\KwOut{$X_g$: Global optimal solution.}
\textbf{Initialize:} Randomly generate $N$ individuals $X$ as $[K, \alpha]$, calculate fitness, and find global optimum $X_g$.\\
\For{$t = 1$ \textbf{to} $max\_iter$}{
    \For{each individual $X_i \in X$}{
        \textit{// Global Search:}\\
        Randomly select $X_j, X_k$, generate $r_1, r_2 \sim U(0, 1)$, update:
        \[
        X_i^{\text{new}} = X_i + r_1 (X_g - X_i) + r_2 (X_j - X_k)
        \]
        \textit{// Local Search:}\\
        \If{$X_i$ is close to $X_g$}{
            Perturb with $\delta \sim U(-0.1, 0.1)$:
            \[
            X_i^{\text{new}} = X_i + \delta \cdot (X_g - X_i)
            \]
        }
        \textit{// Cooperation:}\\
        \If{fitness($X_i$) is worse than local leader $X_{\text{leader}}$}{
            Adjust:
            \[
            X_i^{\text{new}} = X_{\text{leader}} + r_3 \cdot (X_i - X_{\text{leader}})
            \]
        }
        Replace $X_i$ if fitness($X_i^{\text{new}}$) improves.\\
    }
    Update $X_g$ with best fitness in $X$.\\
    \If{convergence criteria met}{
        \textbf{break.}
    }
}
\textbf{Output:} Global optimal solution $X_g$ and its fitness.
\end{algorithm*}

\subsection{BPM Estimation Based on R-peak Detection}

After applying NRBO-VMD to the CMS signal, heart rate estimation is performed using a robust R-peak detection algorithm [7]. The processed signal undergoes dynamic thresholding to identify R-peaks. Temporal constraints ensure physiological plausibility by enforcing a refractory period to eliminate false positives. Heart rate is then computed as the number of detected R-peaks per minute.
\section{Experimental Results}
\subsection{Experimental Setups}

The experimental setups in this experiment involves analyzing the echo signals of cardiac activity captured by FMCW radar from back movements and comparing these signals to the standard ECG measurements. The experiment involved 18 participants, aged between 19 and 26 years. Each participant was measured in a 3-minute session. A 1 minute signal was extracted for each indivudual to ensure a best quality of the signal slicing. Within the 1 minute period, R-peaks were extracted from the ECG data, and the referencing beats per minute (BPM) values were calculated for comparison and validation.

The radar device was positioned 20 cm behind the participant in a seated posture to monitor cardiac mechanical signals. Simultaneously, standard ECG signals were measured by method of three-lead electrodes and recorded to serve as a reference for comparison with the radar-derived heartbeat signals.

The radar system utilized in the experiment was the Texas Instruments IWR6843 ISK with DCA1000C data acquisition board [8], operating at a frequency of 60 GHz with a sampling rate of 2000 Hz. ECG signals were acquired using three electrodes fixed on the left shoulder blade, right shoulder blade the left abdomen of each individual (Fig. 1). 

\subsection{Results and Discussion}
\subsubsection{Analysis of Reconstructed CMS}
Firstly, the CMS data of subject 18 were analyzed, and the frequency domain images optimized using NRBO were compared with those obtained directly using VMD. It was observed that NRBO-VMD yielded a more concentrated frequency distribution, indicating superior signal consistency and quality.

\subsubsection{Comparison of Different Signal Reconstruction Methods}
After conducting a case study involving a total of 18 subjects, the heart rate estimation using the NRBO-VMD method was compared with the BPF, VMD, and GA-optimized VMD methods in terms of performance.

\begin{figure}[h]
\centerline{\includegraphics[width=0.9\columnwidth]{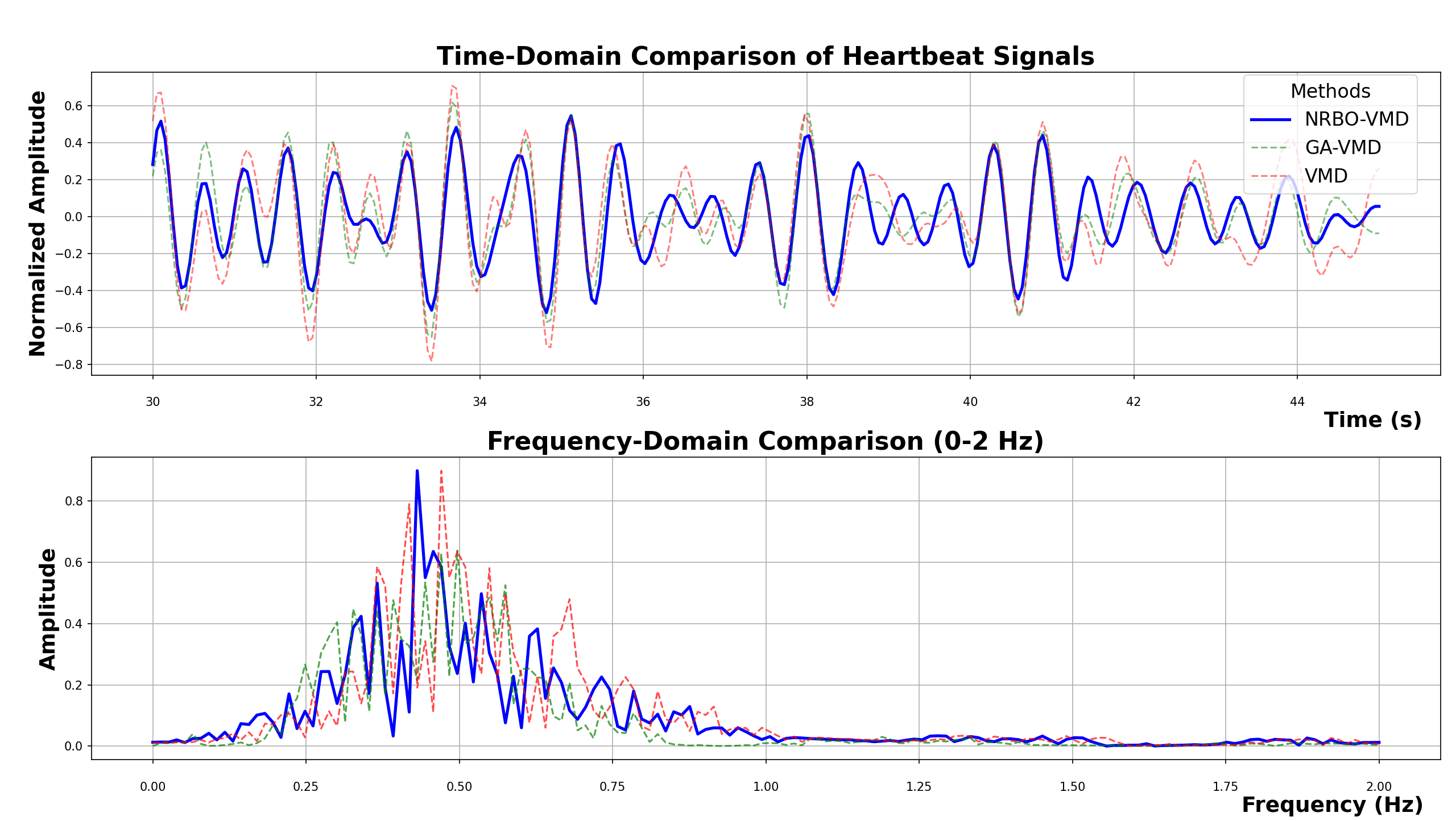}}
\caption{The comparison of  NRBO-VMD, GA-VMD and VMD}
\label{fig}
\end{figure}

\begin{figure}[h]
\centerline{\includegraphics[width=0.9\columnwidth]{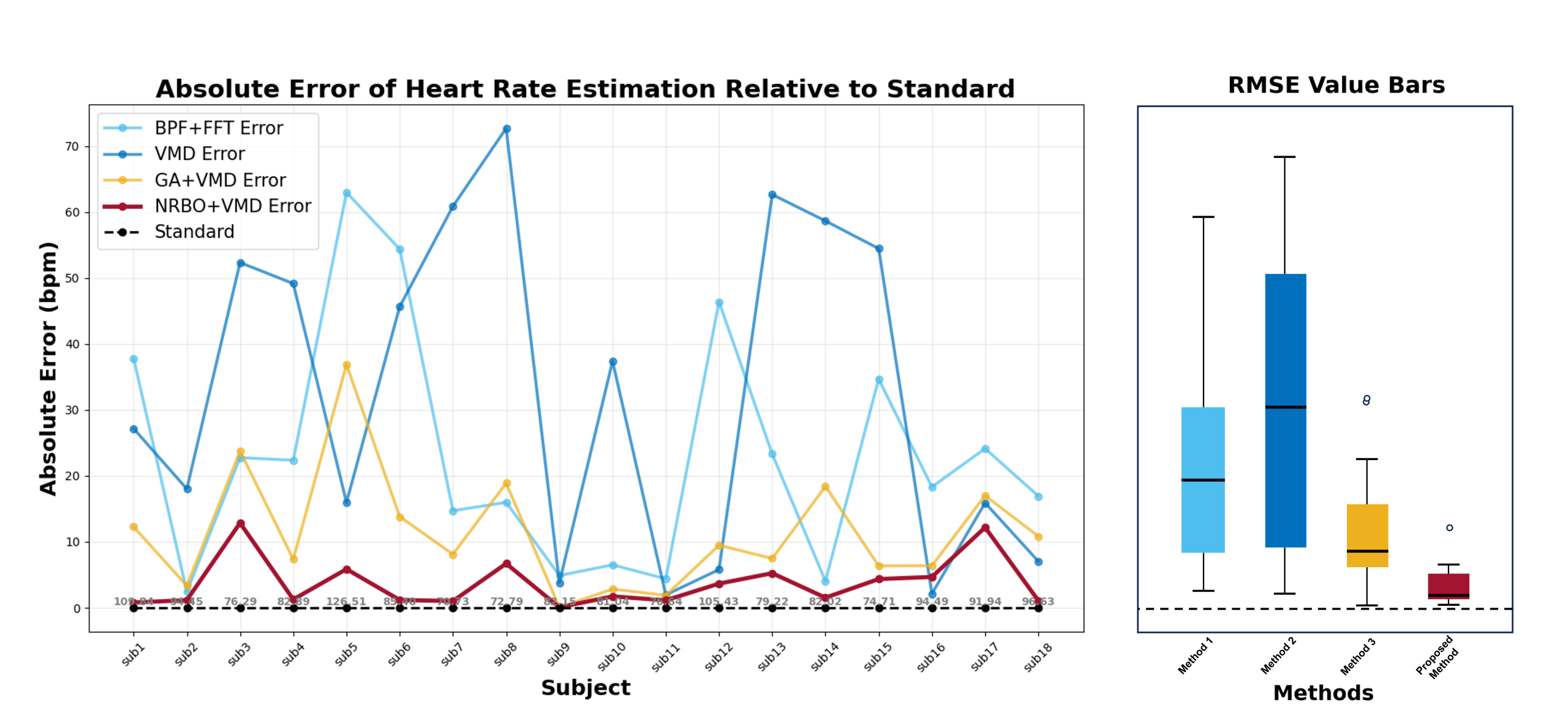}}
\caption{The absolute error of four methods}
\label{fig2}
\end{figure}

\begin{table}[b]
\centering
\caption{Heart Rate Estimation Results}
\begin{tabular}{|c|c|c|c|}
\hline
\textbf{Work} & \textbf{Method} & \textbf{RMSE (bpm)} & \textbf{Accuracy (\%)} \\
\hline
Hu \textit{et al.} [4] & BPF+FFT & 28.924 & 67.115 \\
\hline
Xu \textit{et al.} [2] & VMD & 40.579 & 53.863 \\
\hline
Du \textit{et al.} [3] & GA+VMD & 14.409 & \textcolor{blue}{83.618} \\
\hline
\rowcolor[HTML]{D3D3D3} 
\textbf{Proposed method} & \cellcolor[HTML]{D3D3D3} NRBO+VMD & \cellcolor[HTML]{D3D3D3} \textbf{5.208} & \cellcolor[HTML]{D3D3D3} \textbf{\textcolor{red}{94.078}} \\
\hline
\end{tabular}
\label{tab:heart_rate_results}
\end{table}

The experimental results underscore the superior performance of the proposed NRBO+VMD method compared to baseline approaches, including Bandpass Filtering (BPF), standard VMD, and Genetic Algorithm-optimized VMD (GA+VMD). Specifically, the NRBO+VMD achieved an accuracy rate of 94.078\% and a root-mean-square error (RMSE) of just 5.208 bpm, demonstrating its robustness in extracting accurate heartbeat signals from radar-based CMS. This represents a substantial improvement over traditional techniques, such as GA+VMD, which achieved an accuracy of 83.618\% with an RMSE of 14.409 bpm. These findings validate the efficiency of NRBO+VMD in enhancing the separation of IMFs, which are critical for accurate cardiac signal reconstruction.

\section{Conclusion}

This study presents a novel approach for noncontact heartbeat estimation using a 60 GHz mmWave radar system, combining VMD with a NRBO algorithm. The proposed method outperforms traditional signal denoising and reconstruction techniques, achieving an estimation accuracy of 94.078\% and a RMSE of 5.208 bpm. By optimizing the IMF and penalty selection, this work highlights the potential the application of adaptive signal reconstruction algorithm in heart beat estimation. The findings lay the foundation for future advancements in precise noncontact vital sign monitoring technologies.


\begin{thebibliography}{0}
\providecommand{\natexlab}[1]{#1}
\providecommand{\url}[1]{#1}
\csname url@samestyle\endcsname
\providecommand{\newblock}{\relax}
\providecommand{\bibinfo}[2]{#2}
\providecommand{\BIBentrySTDinterwordspacing}{\spaceskip=0pt\relax}
\providecommand{\BIBentryALTinterwordstretchfactor}{4}
\providecommand{\BIBentryALTinterwordspacing}{\spaceskip=\fontdimen2\font plus
\BIBentryALTinterwordstretchfactor\fontdimen3\font minus \fontdimen4\font\relax}
\providecommand{\BIBforeignlanguage}[2]{{%
\expandafter\ifx\csname l@#1\endcsname\relax
\typeout{** WARNING: IEEEtranN.bst: No hyphenation pattern has been}%
\typeout{** loaded for the language `#1'. Using the pattern for}%
\typeout{** the default language instead.}%
\else
\language=\csname l@#1\endcsname
\fi
#2}}
\providecommand{\BIBdecl}{\relax}
\BIBdecl

\end{thebibliography}


\begin{thebibliography}{00}
\bibitem{b1} W. Hoffmann, A. Lowe, M. Wilson, and M. M. Y. Kuo, "Investigation of Non-contact Electrodes for Electrocardiogram Monitoring," in \textit{Proc. 45th Annu. Int. Conf. IEEE Eng. Med. Biol. Soc. (EMBC)}, Sydney, Australia, 2023, pp. 1-4.
\bibitem{b2} L. Liu, S. Zhang, and W. Xiao, "Non-Contact Vital Signs Detection Using mm-Wave Radar During Random Body Movements," in \textit{Proc. 16th IEEE Conf. Ind. Electron. Appl. (ICIEA)}, Chengdu, China, 2021, pp. 1244-1249.
\bibitem{b3} X. Xu et al., "mmECG: Monitoring Human Cardiac Cycle in Driving Environments Leveraging Millimeter Wave," in \textit{IEEE INFOCOM 2022 - IEEE Conf. Comput. Commun.}, London, United Kingdom, 2022, pp. 90-99, doi: 10.1109/INFOCOM48880.2022.9796912.
\bibitem{b4} Y. Du, A. Yang, B. Li, and F. Zhang, "77GHz Millimeter-Wave Radar Vital Signs Detection Based on GA-VMD Algorithm," in \textit{Proc. 2nd Int. Conf. Artif. Intell., Big Data, Algorithms (CAIBDA)}, Nanjing, China, 2022, pp. 1-7.
\bibitem{b5} Y. Ben-gong and L. Xiao-Jing, "Research and Improvement of Immune Genetic Algorithm," in \textit{Proc. 2nd IEEE Int. Conf. Inf. Manag. Eng.}, Chengdu, China, 2010, pp. 563-566.
\bibitem{b6} Y. Hu, Z. Xia, and F. Xu, "Using FMCW Millimeter-Wave Radar to Realize the Detection of Vital Signs," in \textit{Proc. Int. Conf. Microw. Millim. Wave Technol. (ICMMT)}, 2021, pp. 978-1-6654-3437-9.
\bibitem{b7} Z. L. Xia, X. H. Wang, H. B. Wei, and Y. Xu, "Detection of Vital Signs Based on Variational Mode Decomposition Using FMCW Radar," in \textit{Proc. Int. Conf. Microw. Millim. Wave Technol. (ICMMT)}, 2021, pp. 978-1-6654-3437-9.
\bibitem{b8} C. Fan, D. Chen, L. Li, and J. Cui, "Equiprobable symbolization sample entropy: A complexity measure for discriminating two-phase flow dynamics," in \textit{Proc. 35th Chinese Control Conf. (CCC)}, Chengdu, China, 2016, pp. 1268-1272, doi: 10.1109/ChiCC.2016.7553262.
\bibitem{b9} M. M. Ali, M. K. Hassan, and M. H. Kabir, "Precise Detection and Localization of R-peaks from ECG Signals," \textit{IEEE Access}, vol. 8, pp. 145452-145463, 2020, doi: 10.1109/ACCESS.2020.3012169.
\bibitem{b10} Texas Instruments, "IWR6843ISK: Millimeter-wave sensor," Texas Instruments, [Online]. Available: https://www.ti.com.cn/tool/cn/IWR6843ISK [Accessed: Dec. 14, 2024].
\end{thebibliography}
\end{document}